# DATA-DRIVEN SHOULDER INVERSE KINEMATICS

YoungBeom Kim[1], Byung-Ha Park[1], Kwang-Mo Jung[1], and JungHyun Han[2]

[1]Korea Electronics Technology Institute, Seoul, Korea
[2]Department of Computer Science and Engineering, Korea University, Seoul, Korea

*ABSTRACT*

*This paper proposes a shoulder inverse kinematics (IK) technique. Shoulder complex is comprised of the sternum, clavicle, ribs, scapula, humerus, and four joints. The shoulder complex shows specific motion pattern, such as Scapulo humeral rhythm. As a result, if a motion of the shoulder isgenerated without the knowledge of kinesiology, it will be seen as un-natural. The proposed technique generates motion of the shoulder complex about the orientation of the upper arm by interpolating the measurement data. The shoulder IK method allows novice animators to generate natural shoulder motions easily. As a result, this technique improves the quality of character animation.*

## 1. INTRODUCTION

The natural character animation is important factor for realistic computer graphics applications. However, high quality character animation generation is a time-consuming process. The production pipeline for real-time 3D character animation can be summarized as the *modelling*, *rigging*, *animation*, and *skinning* stages. In the modelling stage, a polygonal character mesh is generated in the rest pose. The stage of rigging comprises three steps. The first step is the insertion of bones into the mesh. The second step is setting bone controllers to allow convenient handling of bone movements, i.e., inverse kinematics (IK). The last step involves binding bone weights that represent the influence of the bones on a vertex. After the rigging stage, motion of a character is created using key-frames in the animation stage. These stages are done offline. In the skinning stage, a skinning technique such as linear blend skinning (LBS) produces the real-time character animation.

The shoulder inverse kinematics (Shoulder IK) is involved in the second step of the rigging stages. The setting of bone controllers normally depends on the artist's knowledge and experience. While attempts have been made to automate this stage, it is still challenging to automate the rigging of the shoulder region because of the complex structure and movement pattern of the shoulder. For example, the scapulohumeral rhythm is shown during the abduction of the shoulder : when the shoulder abducts more than 30 degrees, abduction of the GH joint and upward rotation of the sternoclavicular joint contributes to shoulder abduction at a ratio of 2:1.Without understanding of the kinesiological knowledge, natural shoulder motion is difficult to be generated.

The shoulder IK generates motion of the shoulder complex about the orientation of the upper arm by interpolating the measurement data. The proposed method also lowers the degrees of freedom in the shoulder complex. As a result, the shoulder IK enables novice artists to efficiently create plausible shoulder motions without an understanding of the shoulder movements. Further the technique can be used in real-time computer graphics applications, such as VR contents and video games.

                                                      

## 2. RELATED WORK

The related works are categorized into two topics: *inverse kinematics* and *shoulder motion*

### 2.1. Inverse Kinematics

IK enables animators to generate motions of a character easily. The character has the hierarchy which starts from a root to limbs such as arms and legs. To produce the full body motion with forward kinematics, an animator should assign the transformation of the whole joints. However, IK simplifies this task, by computing the joints' angle from the end-effector's position.

IK can be classified into the four categories. The first category is analytical methods. They have closed form solutions. The methods proposed by Wu *et al.*[1] and Gan*etal.*[2] set the end-effectors' position and solve the IK problem by studying possible relative posture properties. These methods are fast and accurate, but have limits on the application area.

The second category is a numerical method based approach. With IK in this cluster, the better animation can be achieved, but normally requires a computational burden, which arises from the iteration. Many methods in this category are based on the Jacobian matrix [3][4][5]. The Jacobian matrix represents the first order partial derivatives, which converts the changes of the local rotation into the end-effectors' movement. Therefore the IK can be modelled as the error minimization problem with the inverse of the Jacobian matrix. Several methods have been proposed for computation of the inverse of the Jacobian, such as Pseudo-Inverse Jacobian[6] and Jacobian Transpose [4].

The third category is heuristics based approach. Cyclic Coordinate Descent (CCD) [7] is one of the most popular methods. The main idea is to align the joint with the end effector and the target at a time, and to bring the end effector closer to the target. The several extensions have been made, such as [8]. Even though CCD based methods require the iteration process, they are extremely efficient, because they do not require any matrix computation.

The last category is machine learning based approach. Real-time control of the end effector of a humanoid robot in external coordinates requires computationally efficient solutions of the inverse kinematics problem. For this purpose, the locally weighted projection regression, support vector machine, artificial neural networks are adapted [9][10][11]. The shoulder IK falls into this cluster. It uses the measurement data of shoulder movement and reproduces the unmeasured data by using the blending technique.

### 2.2. Shoulder motion

Biomechanics researchers have argued to find the pattern between the movement of the humerus and the other bones in the shoulders, i.e. clavicle and scapula. The *scapulohumeralrhythm*[12] is one of the most popular models. The scapulohumeral rhythm is detected in the 2D view and it explains shoulder movements during abduction as follows: when the shoulder abducts more than 30 degrees, abduction of the GH joint and upward rotation of the SC joint contribute to shoulder abduction at a ratio of 2:1.

Recently, there are many approaches to finding the pattern of the shoulder motion, i.e., *shoulder rhythm* in 3-dimensional space. de Groot and Brand [13] measured the right arm at 23 positions, spread over four vertical planes of elevation: 30, 60, 90, 120 degrees with the frontal plane, and six levels of elevation at each plane: 0, 30, 60, 90, 120, 150 degrees. The regression model predicts the motion of the clavicle and the scapula by five linear regression equations. Xu*etal.*[14] also present the regression model predicts the motion of the clavicle and the scapula about the movements of the humerus. However, they also have sparse measurements of the elevation: 0, 30, 60, 90, 120, 150 degrees and reports the existence of the error between measured and predicted joint angles. The shoulder IK uses the more fine data of [15] for realistic animation of the shoulder complex.

## 3. SHOULDER COMPLEX

The shoulder complex comprises of the sternum, clavicle, ribs, scapula, humerus, and four joints, as shown in Figure 1. The first joint, the sternoclavicular (SC) joint, connects the sternum and the clavicle. At the lateral end of the clavicle, the acromioclavicular (AC) joint attaches the scapula to the clavicle. The interface between the scapula and thorax is a non-anatomic joint called scapulothoracic (ST) joint. The most distal joint is the glenohumeral (GH) joint, which connects the scapula and the humerus. In this paper, the movement of ST joint is described as a combined motion of the SC and AC joints. Note that this paper considers the sternum and ribs as the fixed objects.

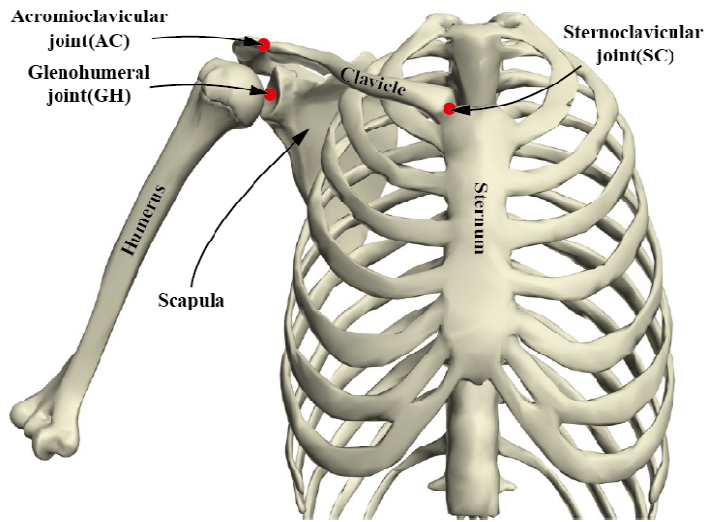

Figure 1.The bones and joints of the right shoulder complex
.

In physiology, for abbreviated description of the shoulder motion, the sagittal, scapular, and frontal planes are introduced. Figure 2shows the planes.

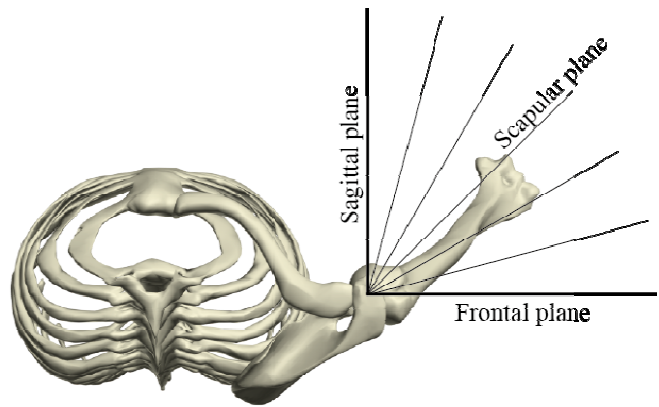

Figure 2.The planes for the abbreviate description of shoulder motion.

To explain the motion of each bone, the coordinates should be defined. International Society of Biomechanics (ISB) presents the recommendation on definitions of coordinate systems for the reporting of human joint motion [16]. The coordinate systems are defined by using the anatomical landmarks, which represent the specific locations on the bones or joints. The detailed description of the landmarks is described in the work of the Wu *et al.*[16]. However, there are conflicts for the definition of the *correct* coordinates and some researchers modify the definitions. An example can be found in Ludewig*et al.*[15].

In this paper, the coordinates follow the work of Ludewig*et al.* [17], which generally follows Wu *et al.*[16]. The coordinates of thorax $x_t y_t z_t$ are aligned with the sagittal, transverse and frontal planes and the origin is at the IJ landmark, as shown in Figure 3-(a).

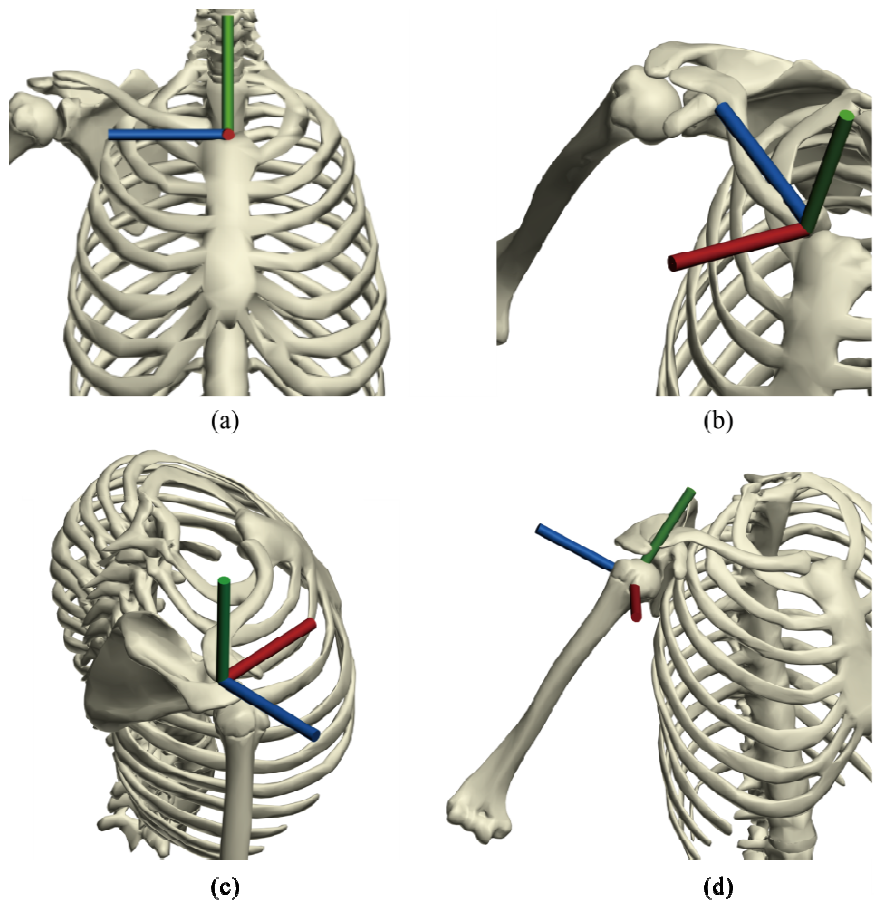

(a) (b)

(c) (d)

Figure 3.The coordinate systems: (a) thorax, (b) clavicle, (c) scapula and (d) humerus. The red, green, and blue lines represent the *x*, *y*, and *z*-axes, respectively.

The coordinates of clavicle $x_c y_c z_c$ are set as follows: the $z_c$-axis is directed along its long axis from the SC landmark to the AC landmark. The $x_c$-axis is the line perpendicular to the $z_c$-axis and $y_t$-axis, pointing forward. The $y_c$-axis is the common line perpendicular to the $x_c$ and $z_c$-axes pointing upward. In addition, the origin of the clavicle's coordinates is at the SC landmark. Figure 3-(b) shows the clavicle's coordinates.

For the scapula's coordinates $x_s y_s z_s$, the new land mark located at the posterior aspect of the acromioclavicular joint is defined. It is called PAC landmark. In this paper, by following

Ludewig *et al.*[17], the PAC landmark is used instead of the AC landmark. As a result, the $z_s$-axis is directed from the TS to the PAC landmarks. The $x_s$-axis is the pointing forward line perpendicular to the plane formed by the AI, PAC, and TS landmarks. The $y_s$-axis is the line perpendicular to the $x_s$ and $z_s$-axes, pointing upward. The origin of the coordinates is at the PAC landmark. Figure 3-(c) shows the scapula's coordinates.

The humerus coordinate system $x_h y_h z_h$ is set as follows: the $z_h$-axis is directed parallel to a line connecting the EM and EL landmarks. The $y_h$-axis is aligned along the shaft and the $x_h$ is the line perpendicular to the $y_h$ and $z_h$, pointing to the right. The origin of the coordinates is at the GH landmark. Figure 3-(d) shows the clavicle's coordinates.

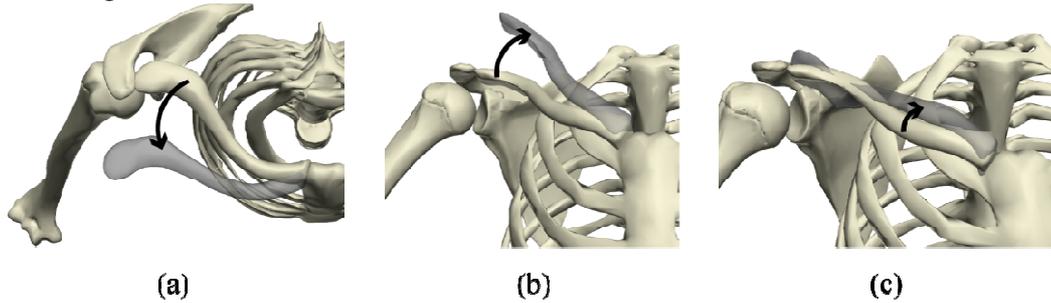

Figure 4. Motions of the clavicle: (a) protraction in the superior view of a right shoulder, (b) elevation in the anterior view of a right shoulder, and (c) posterior rotation in the lateral view.

With the coordinates, the motion of each bone can be described. The motion of the clavicle is comprised of protraction, elevation, and posterior rotations as shown in Figure 4.

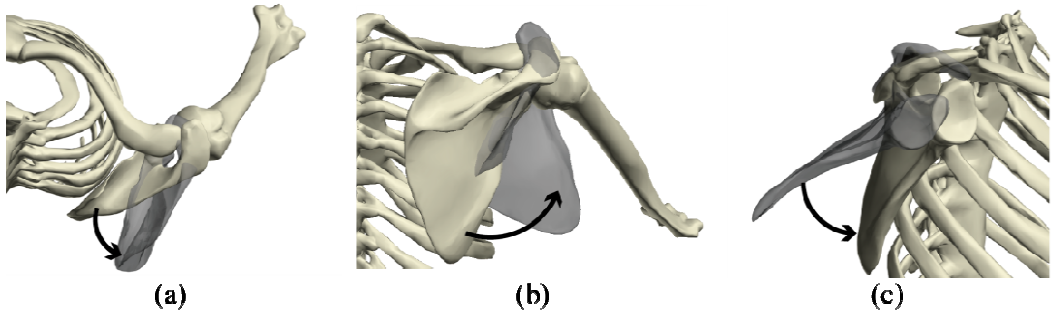

Figure 5. Motions of the scapula: (a) internal rotation in the superior view, (b) upward rotation in the posterior view, and (c) posterior tilting in the lateral.

The motions of the scapula are composed of internal rotation, upward rotation, and posterior tilting as shown in Figure 5. The motions of the humerus are comprised of humeral elevation, the plane of elevation, and axial rotation as shown in Figure 6.

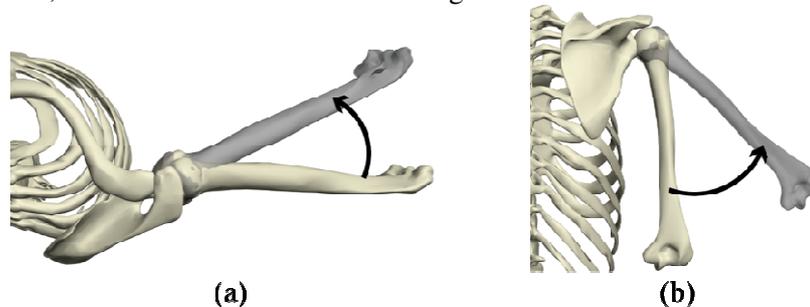

Figure 6. Motions of the humerus: (a) internal rotation in the superior view, and (b) upward rotation in the posterior view.

## 4. SHOULDER IK METHOD

The shoulder IK method enables animators to generate the realistic shoulder motion. Generating plausible shoulder motions without knowledge about the shoulder rhythm, which describes the pattern of shoulder motion, is not a trivial work. Figure 7 overviews this method. The shoulder IK generates the realistic motions of the SC, AC, and GH joints from the humerothoracic elevation angle $\theta$, which represents the humeral elevation angle relative to the thorax and the orientation angle $\psi$ of the elevation plane. The final output is computed by interpolating the data of the shoulder motion database, which is described in the next subsection. Note that the orientation angle $\psi$ refers to the rotation degree of the frontal plane about the vertical axis.

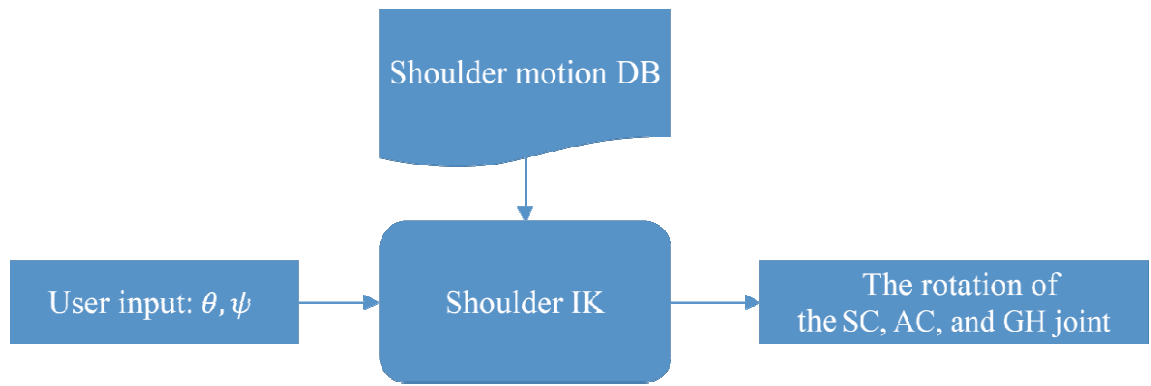

Figure 7.Overview of the Shoulder IK method.

### 4.1. Shoulder Motion Database

The shoulder motion DB contains the measurement data about the shoulder motion. This paper uses the measurement data of Ludewig*et al.*[17]. They measured the motion of the shoulder bones by using invasive method, which inserts the bone pins. Since the instrumentation is in direct contact with the shoulder bones, invasive methods are assumed to be *gold standards*.

The measurement data are comprised of the Euler angles of the SC joint ($\hat{E}^{SC}$), AC joint ($\hat{E}^{AC}$), and GH joints ($\hat{E}^{GH}$). The Euler angles for each of the SC, AC, and GH joint motions were extracted at 15 degrees of humeral elevation, and for each 5 degree increment of humeral elevation in each of the planes of elevation, i.e., the frontal, scapular, and sagittal planes. The maximum measurement degree of the humeral elevation is 120 degrees. As a result, each joint has rotation measurement data about the 66 (22×3) positions.

For abbreviated description of shoulder motion, the frontal, scapular, and sagittal planes are represented by the orientation angle degrees $\hat{\psi}_0 = 0$, $\hat{\psi}_1 = 40$ and $\hat{\psi}_2 = 90$, respectively. The measured humeral elevation degrees are represented by $\hat{\theta}_0 = 15$, $\hat{\theta}_1 = 20$,..., and $\hat{\theta}_{21} = 120$. Note that $\hat{\psi}$ and $\hat{\theta}$ represent the *measured* positions. As a result, the Euler angles of SC joint about $\theta = 20$ and $\psi = 40$ are represented by $E^{SC}_{\hat{\theta}_1 \hat{\psi}_1}$.

### 4.2. Spherical Interpolation

In general, spherical line a interpolation (slerp) is used for interpolating the rotations:

$$slerp(\mathbf{q}_0, \mathbf{q}_1, t) = \frac{\sin((1-t)\Omega)}{\sin(\Omega)}\mathbf{q}_0 + \frac{\sin(t\Omega)}{\sin(\Omega)}\mathbf{q}_1 \quad (1)$$

where $\mathbf{q}_0$ and $\mathbf{q}_1$ are the rotations, which will be interpolated, while $\Omega$ is the rotation angle between $\mathbf{q}_0$ and $\mathbf{q}_1$. The interpolation weight $t$ is in the range of [0, 1].

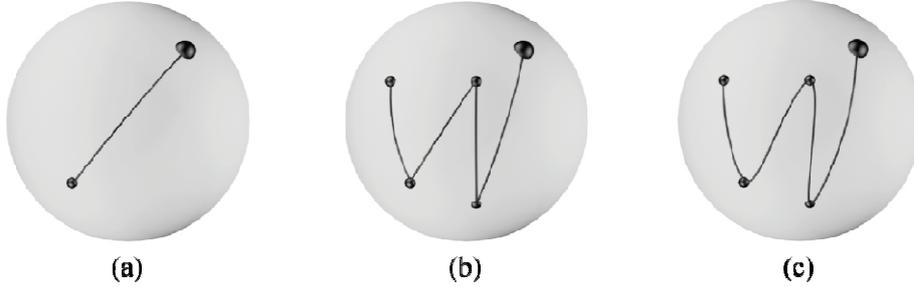

Figure 8.Rotation interpolation results: (a) slerp between two rotations, (b) slerp between a series of rotations, and (c) squad between the series of rotations.

Slerp generates the optimal interpolation between two rotations, but it reveals a problem when interpolating a series of rotations [18]. The quaternion is in the four dimensional space and it cannot be directly visualized. However, a unit quaternion, which represents the rotation, can be visualized by using the unit sphere as shown in Figure 8. In Figure 8, the biggest grey-colored sphere represents the three-dimensional unit sphere. The first key frame is represented as a bigger dot. The other key frames are represented as smaller dots, and interpolated frames are represented as lines, which are called interpolation curves. Figure 8-(a) shows the proper interpolation results of slerp for the two rotations. However, slerp does not generate smooth interpolation curve for the series of rotations as shown as Figure 8-(b).

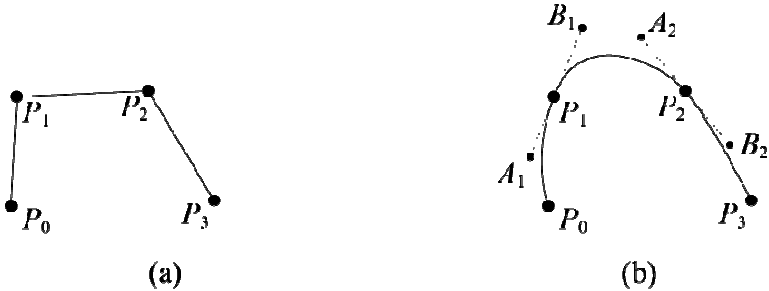

Figure 9.The interpolation results: (a) linear interpolation between a series of points, and (b) the Bézier curve for a series of points. Note that the curve is defined as a third-order curve, where the tangent in the control points is defined by auxiliary points, e.g., the tangent in $P_1$ is defined by the auxiliary points $A_1$ and $B_1$

The same problem is also found in the simple Euclidean space, as shown in Figure 9, and the Bézier curve can resolve the problem, as Figure 9-(b) shows. The Bézier curve from Figure 9-(b) which interpolates the control points $P_1$ and $P_2$ can be expressed as three steps of linear interpolation [18]:

$$lin(x_0, x_1, t) = x_0(1-t) + x_1 t \quad (2)$$

$$Bézier(P_1, P_2, B_1, A_2, t) = lin(lin(P_1, P_2, t), lin(B_1, A_2, t), 2t(1-t)) \quad (3)$$

The similar interpolation for the spherical coordinates, i.e., squad (**s**pherical and **quad**rangle) was presented by Shoe make[19], and it is defined as follows:

$$squad(\mathbf{q}_i, \mathbf{q}_{i+1}, t) = slerp(slerp(\mathbf{q}_i, \mathbf{q}_{i+1}, t), slerp(\mathbf{s}_i, \mathbf{s}_{i+1}, t), 2t(1-t)) \quad (4)$$

$$s_i = q_i \exp\left(-\frac{\log(q_i^{-1}q_{i+1}) + \log(q_i^{-1}q_{i-1})}{4}\right) \quad (5)$$

The equations show that the squad uses the spherical linear interpolation instead of the linear interpolation and $B_1$ and $A_2$ are written as $s_i$ and $s_{i+1}$. The derivation of equation for $s_i$ is out of the focus of this paper. The details are described in the work of Dam *et al.*[18].

### 4.3. IK Based on Spherical Interpolation

The shoulder IK method generates the motion of the shoulder joint from the humerothoracic elevation angle $\theta$ and the orientation angle $\psi$ of the elevation plane by applying the bi-spline interpolation to the shoulder motion measurement data $\hat{E}^{SC}$, $\hat{E}^{AC}$, and $\hat{E}^{GH}$. Figure 10 shows the procedure of the shoulder IK.

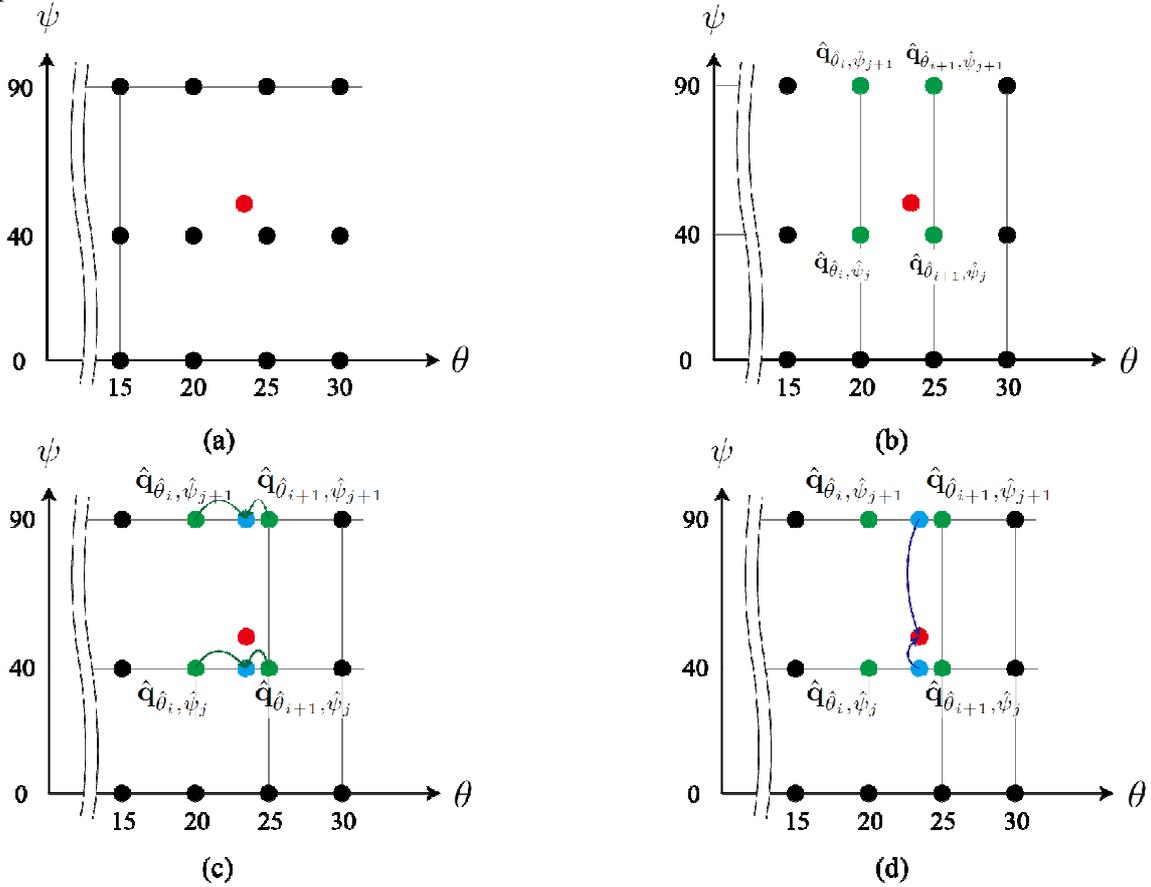

Figure 10. The procedure of the bi-spline interpolation. (a) Input is represented as the red dot. (b) The interpolation candidates are determined and are represented as the green dots. (c) First interpolation about $\theta$ is applied. (d) Final rotation values are calculated by second interpolation.

The first step in the shoulder IK method is conversion of the measurement data $\hat{E}^{SC}$, $\hat{E}^{AC}$, and $\hat{E}^{GH}$ into quaternions $\hat{q}^{SC}$, $\hat{q}^{AC}$, and $\hat{q}^{GH}$, respectively. The converted data are shown as the black dots in Figure 10-(a). It is an offline process. During the online process, the shoulder IK is called each frame with the input which is comprised of $\theta$ and $\psi$. It is shown as the red dot in Figure 10-(a). Then the shoulder IK method determines the four interpolation candidates, depicted as the green dots in Figure 10-(b). The interpolation candidates are represented as $\hat{q}_{\hat{\theta}_i,\hat{\psi}_j}$, $\hat{q}_{\hat{\theta}_i,\hat{\psi}_{j+1}}$, $\hat{q}_{\hat{\theta}_{i+1},\hat{\psi}_j}$, and $\hat{q}_{\hat{\theta}_{i+1},\hat{\psi}_{j+1}}$. As a result, this paper chooses $i$ and $j$ which satisfy the following equations:

$$\hat{\theta}_i < \theta \leq \hat{\theta}_{i+1} \tag{6}$$

$$\hat{\psi}_i < \psi \leq \hat{\psi}_{i+1} \tag{7}$$

When *i* and *j* are computed, the interpolation weights are determined as follows:

$$t_\theta = (\theta - \hat{\theta}_i)/5 \tag{8}$$

$$t_\psi = \begin{cases} \dfrac{\psi - \hat{\psi}_j}{40}, & if\ \psi \leq 40, \\ \dfrac{\psi - \hat{\psi}_j}{50}, & otherwise \end{cases} \tag{9}$$

For example, using the inputs $\theta = 23$ and $\psi = 50$ as in Figure 10, the interpolation candidates and the weight are selected as follows: $\hat{\theta}_i = 20$, $\hat{\theta}_{i+1} = 25$, $\hat{\psi}_j = 40$, $\hat{\psi}_{j+1} = 90$, $t_\theta = 3/5$, and $t_\psi = 10/50$.

The shoulder IK method computes the final rotation $\mathbf{q}'_{\theta,\psi}$ by applying spherical interpolations as shown in Figure 10-(c) and -(d). At first, the quaternion candidates are interpolated about $\theta$ as shown in Figure 10-(c). This process proceeds with the following equations:

$$\mathbf{q}'_{\psi_j} = Squad(\mathbf{q}_{\theta_i,\psi_j}, \mathbf{q}_{\theta_{i+1},\psi_j}, t_\theta) \tag{10}$$

$$\mathbf{q}'_{\psi_{j+1}} = Squad(\mathbf{q}_{\theta_i,\psi_{j+1}}, \mathbf{q}_{\theta_{i+1},\psi_{j+1}}, t_\theta) \tag{11}$$

The final rotation $\mathbf{q}'_{\theta,\psi}$ is calculated by the second interpolation which uses $\mathbf{q}'_{\psi_j}$, $\mathbf{q}'_{\psi_{j+1}}$ and $t_\psi$ as follows:

$$\mathbf{q}'_{\theta,\psi} = Squad(\mathbf{q}'_{\psi_j}, \mathbf{q}'_{\psi_{j+1}}, t_\psi) \tag{12}$$

Figure 10-(d) shows this process.

## 5. RESULT

This section presents experimental results of the shoulder IK. The linear blend skinning method is used for Figure 14.

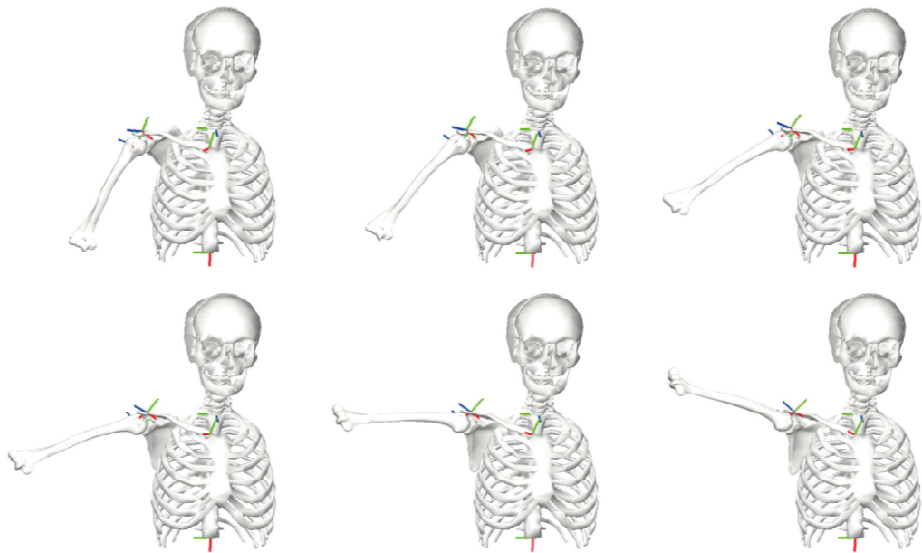

Figure 11.Shoulder inverse kinematics results during abduction in the front view.

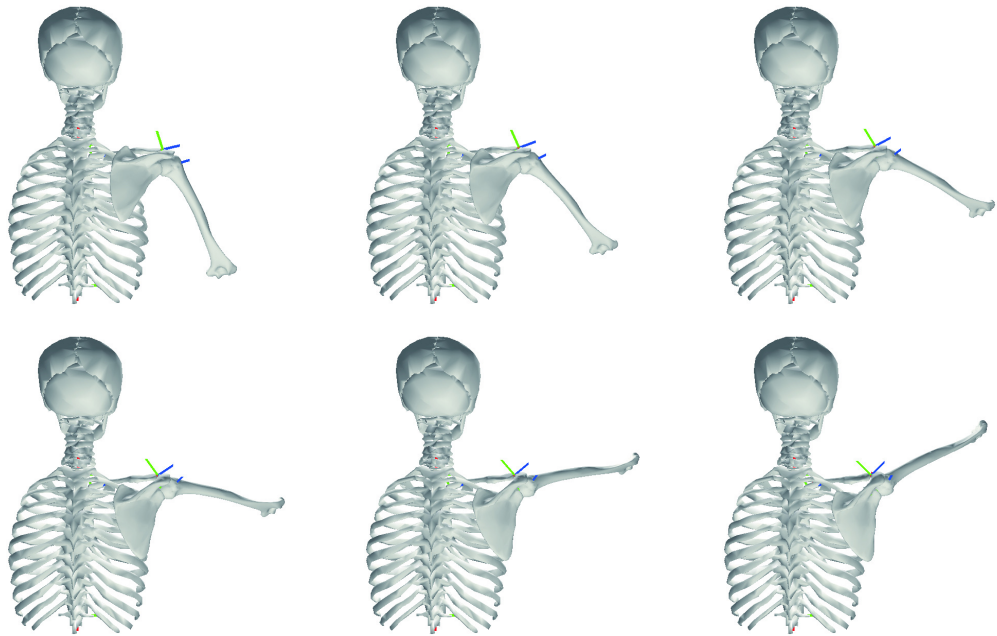

Figure 12.Shoulder inverse kinematics results during abduction in the back view.

The results obtained by using the shoulder IK method are depicted in Figure 11 and 12. They depict the natural motions of the SC, AC, and GH joints during abduction. Figure 13depicts the trajectories of the elbow using the squad-based shoulder IK method and the slerp-based shoulder IK method during the protraction of the arm. It shows that the squad-based shoulder IK generates a much smoother shoulder motion than the slerp-based shoulder IK method.

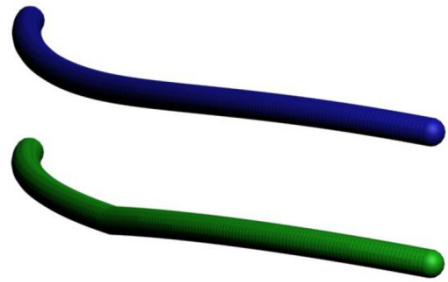

Figure 13.Trajectories of the elbow during arm protraction. The blue and green trajectories are the results obtained using the shoulder IK method with squad and shoulder IK with slerp, respectively.

The results obtained using the shoulder IK methods are compared with the shoulder animation results produced by a novice artist. Figure 14 shows the results. It is evident that the animation produced without using the shoulder IK method contains unnatural skinning near the AC joint.

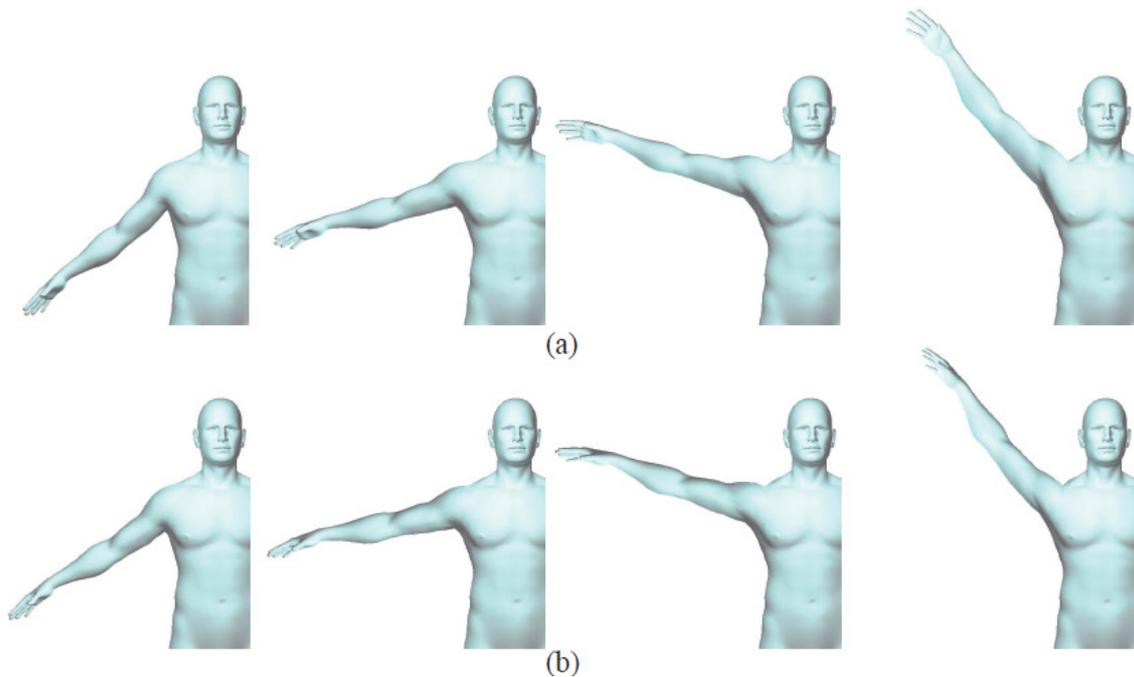

Figure 14.The comparison of the shoulder animations. (a) shows the shoulder IK applied animation and (b) shows the *novice* artist's animation, which only contains the GH joint's rotation.

## 6. CONCLUSION AND FUTURE WORK

The shoulder IK generates plausible shoulder motion by using the measurement data. With this technique, novice artists generate the plausible shoulder animation easily as can be seen in the result section of this technique.

However, the shoulder IK considers only the shoulder part, and therefore its application area is limited. For generalization, this method should be extended to the upper-body or full-body IK. In addition, the comparison between the shoulder IK and the shoulder rhythm should be made. It is worth to apply the proposed method to the biological simulation of characters. The future work will be carried out along these directions.

## ACKNOWLEDGEMENTS

This work was supported by the Ministry of Trade, Industry & Energy grant funded by the Korea government (No. 10051089).

## REFERENCES

[1] X. Wu, L. Ma, Z. Chen, and Y. Gao, "A 12-dof analytic inverse kinematics solver for human motion control," Journal of Information and Computational Science, vol. 1, pp. 137–141, 2004.

[2] J. Q. Gan, E. Oyama, E. M. Rosales, and H. Hu, "A complete analytical solution to the inverse kinematics of the pioneer 2 robotic arm," Robotica, vol. 23, no. 1, pp. 123–129, Jan. 2005.

[3] C. W. Wampler, "Manipulator inverse kinematic solutions based onvectorformulationsanddampedleast-squaresmethods,"IEEETransactions on Systems, Man, and Cybernetics, vol. 16, no. 1, pp. 93–101, Jan 1986.

[4] W. A. Wolovich and H. Elliott, "A computational technique for inverse kinematics," in Decision and Control, 1984. The 23rd IEEE Conference on, Dec 1984, pp. 1359–1363.

[5] J. Yuan, "Local svd inverse of robot jacobians," Robotica, vol. 19, no. 1, pp. 79–86, Jan. 2001.[Online]. Available: http://dx.doi.org/10.1017/S0263574700002769

[6] R. G. Roberts and A. A. Maciejewski, "Singularities, stable surfaces, and the repeatable behavior of kinematically redundant manipulators," The International Journal of Robotics Research, vol. 13, no. 1, pp. 70–81, 1994. [Online]. Available: http://ijr.sagepub.com/content/13/1/70.abstract.

[7] L. C. T. Wang and C. C. Chen, "A combined optimization method for solving the inverse kinematics problems of mechanical manipulators," IEEE Transactions on Robotics and Automation, vol. 7, no. 4, pp. 489– 499, Aug 1991.

[8] A. A. Canutescu and R. L. Dunbrack, "Cyclic coordinate descent: A robotics algorithm for protein loop closure," Protein Sci., vol. 12, no. 5, pp. 963–972, May 2003.

[9] A. D'Souza, S. Vijayakumar, and S. Schaal, "Learning inverse kinematics," in Intelligent Robots and Systems, 2001.Proceedings. 2001 IEEE/RSJ International Conference on, vol. 1, 2001, pp. 298–303 vol.1.

[10] E. Sariyildiz, K. Ucak, G. Oke, H. Temeltas, and K. Ohnishi, "Support vector regression based inverse kinematic modeling for a 7-dof redundant robot arm," in Innovations in Intelligent Systems and Applications (INISTA), 2012 International Symposium on, July 2012, pp. 1–5.

[11] A. Morell, M. Tarokh, and L. Acosta, "Inverse kinematics solutions for serial robots using support vector regression," in Robotics and Automation (ICRA), 2013 IEEE International Conference on, May 2013, pp. 4203–4208.

[12] V. T. Inman, J. B. deC. M. Saunders, and L. C. Abbott, "Observations on the function of the shoulder joint," The Journal of Bone & Joint Surgery, vol. 26, no. 1, pp. 1–30, 1944.

[13] J. H. de Groot and R. Brand, "A three-dimensional regression model of the shoulder rhythm," ClinBiomech (Bristol, Avon), vol. 16, no. 9, pp. 735–743, Nov 2001.

[14] X. Xu, J. H. Lin, and R. W. McGorry, "A regression-based 3-D shoulder rhythm," J Biomech, vol. 47, no. 5, pp. 1206–1210, Mar 2014.

[15] P. M. Ludewig, D. R. Hassett, R. F. LaPrade, P. R. Camargo, and J. P. Braman, "Comparison of scapular local coordinate systems," Clinical Biomechanics, vol. 25, no. 5, pp. 415 – 421, 2010. [Online]. Available: http://www.sciencedirect.com/science/article/pii/S0268003310000331

[16] G. Wu, F. C. van der Helm, H. E. Veeger, M. Makhsous, P. Van Roy, C. Anglin, J. Nagels, A. R. Karduna, K. McQuade, X. Wang, F. W. Werner, and B. Buchholz, "ISB recommendation on definitions of joint coordinate systems of various joints for the reporting of human joint motion–Part II: shoulder, elbow, wrist and hand," J Biomech, vol. 38, no. 5, pp. 981–992, May 2005.

[17] P. M. Ludewig, V. Phadke, J. P. Braman, D. R. Hassett, C. J. Cieminski, and R. F. LaPrade, "Motion of the shoulder complex during multiplanar humeral elevation," J Bone Joint Surg Am, vol. 91, no. 2, pp. 378–389, Feb 2009.

[18] E. B. Dam, M. Koch, and M. Lillholm, "Quaternions, interpolation and animation," Tech. Rep., 1998.

[19] K. Shoemake, "Quaternion calculus and fast animation," in ACM SIGGRAPH Courses Notes, 1987

**Authors**

**Young Beom Kim** is an associate research engineer of the Korea Electronics Technology Institute (KETI). He received a BS(2008) degree in Computer Science and Engineering at Korea University. He also received MS(2010) and PhD(2016) degrees in Computer Science and Engineering at same university. His research interests focus on real-time character animation and virtual training.

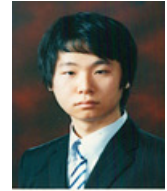

**Byung-Ha Park** is a senior research engineer of the Korea Electronics Technology Institute (KETI). He received a BS(1999) degree in Computer Science at Sejong University, obtained an MS(2001) degree in Computer Engineering at Sejong University. His research interests focus on smart media contents, realistic media, and virtual training.

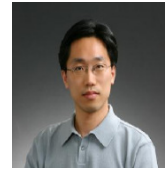

**Kwang-Mo Jung** is a principle research engineer of the Korea Electronics Technology Institute (KETI). He received a BS(1990) degree in Electronics and Communication at KwangWoon University. He also received MS (2002) and PhD (2006) degrees in Electronics and Communication at same university. His research interests focus on smart media contents, realistic media, and virtual training.

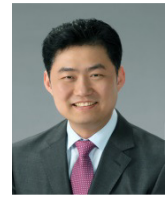

**JungHyun Han** is a professor in the Department of Computer Science and Engineering at Korea University, where he directs the Interactive3D Media Laboratory. Prior to joining Korea University, he worked at the School of Information and Communications Engineering of Sungkyunkwan University, in Korea, and at the Manufacturing Systems Integration Division of the US Department of Commerce National Institute of Standards and Technology (NIST). He received a BS degree in Computer Engineering at Seoul National University, an MS degree in Computer Science at the University of Cincinnati, and a PhD degree in Computer Science at USC.

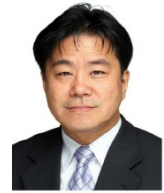